\documentclass[final,5p,times,twocolumn,authoryear]{elsarticle}

\usepackage{amsmath, amssymb, amsfonts, amsthm, fouriernc, mathtools}
\usepackage{gensymb} % use the \degree symbol
\usepackage{float}
\graphicspath{{figures/}}
\usepackage[]{units}
\usepackage{caption}
\usepackage{subcaption}
\usepackage[usenames, dvipsnames]{color} % use colored text with \color{red}{text}%
\usepackage{multirow}
\usepackage[switch]{lineno} % add line numbering

\usepackage{xr}
\journal{Energy and Buildings}

\begin{document}
%\linenumbers

\begin{frontmatter}

%% Title, authors and addresses

\title{Estimating country-specific space heating threshold temperatures from national consumption data}

\author[rvt,rvt2]{S. ~Kozarcanin\corref{cor1}}
\ead{sko@eng.au.dk}

\author[rvt]{G. B. ~Andresen}
\ead{gba@eng.au.dk}

\author[rvt2]{I. ~Staffell}
\ead{i.staffell@imperial.ac.uk}

\cortext[cor1]{Corresponding author}
%\cortext[cor2]{Principal corresponding author}

\address[rvt]{Department of Engineering, Aarhus University, Inge Lehmanns Gade 10, 8000 Aarhus, Denmark}
\address[rvt2]{Centre for Environmental Policy, Imperial College London, 16 Princes Gardens, SW7 1NE London, UK}

\address{}

\begin{abstract}
\noindent Space heating in buildings is becoming a key element of sector-coupled energy system research. Data availability limits efforts to model the buildings sector, because heat consumption is not directly metered in most countries. Space heating is often related to weather through the proxy of heating degree-days using a specific heating threshold temperature, but methods vary between studies. This study estimates country-specific heating threshold temperatures using widely and publicly available consumption and weather data. This allows for national climate and culture-specific human behaviour to be captured in energy systems modelling. National electricity and gas consumption data are related to degree-days through linear models, and Akaike's Information Criteria is used to define the summer season in each country, when space heating is not required. We find that the heating threshold temperatures computed using daily, weekly and monthly aggregated consumption data are statistically indifferent. In general, threshold temperatures for gas heating centre around 15.0 $\pm$ 1.7 $\degree$C (daily averaged temperature), while heating by electricity averages to 13.4 $\pm$ 2.4 $\degree$C. We find no evidence of space heating during June, July and August, even if heating degree-days are present. %These results suggest that current estimations of space heating might be under or overestimated severely for some countries. 
\end{abstract}


\end{frontmatter}

\section{Introduction}

\noindent Two thirds of the energy consumed in north European homes is for space heating, compared to just under a third in the US and China \citep{IEA_OECD}. In 2015, the European heating sector accounted for more than 50\% of the final energy demand of 6110 TWh/yr \citep{HRE4}. Together, the production of electricity and heat accounted for approximately 30\% of  total CO$_2$ emissions, with heat production accounting for more than half of this share \citep{iea2}. Decarbonizing the energy sector, and space heating in particular, is therefore central in limiting global warming. Former studies such as  \cite{kozarcanin2018climate} or \cite{schaeffer2012energy} have shown that the combined impact of climate change on weather-dependent electricity generation and demand is negligible. \cite{kozarcanin2018climate} shows further that most key properties of large-scale renewable-based electricity system are robust against climate change. The electricity sector is therefore already being decarbonized, most efficiently by increasing the share of renewables. However, heat does not have the same rate of technology innovation, clean options are not reducing rapidly in cost \citep{staffell2012review, staffell2018role}, and so progress is very slow \citep{uk_2018}. Natural gas, fuel oil and coal fired boilers are the main source of heat production for the majority of European countries \citep{fleiter2016mapping}, and relatively few countries (primarily the Nordic countries) have a significant share of lower-carbon options. \\

%-------------
\begin{figure*}
    \centering
        \includegraphics[width=0.99\textwidth]{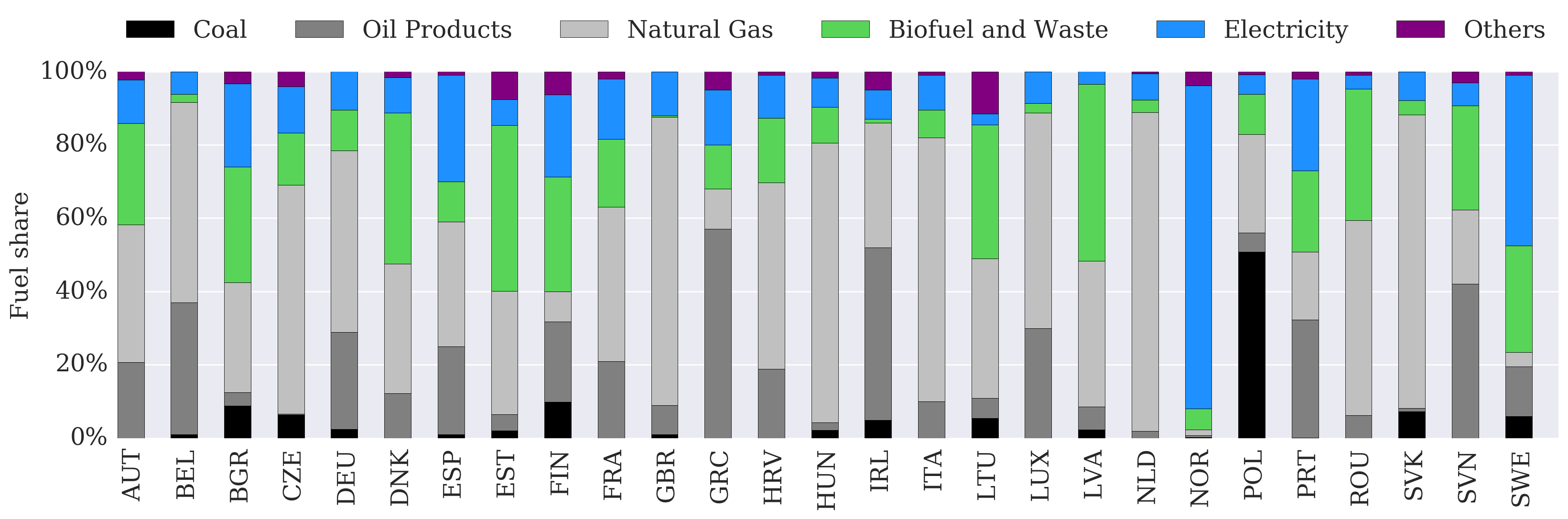}
    \caption{Fuel shares of the final energy demand for the countries included in this study. Based on data compiled from \cite{stratego1}, \cite{vivid} and \citep{european_comission}. Switzerland (CHE) is excluded due to missing data. Countries are referred to by their three-letter ISO codes.} 
    \label{fig: fuel_share}
\end{figure*}
%-------------

\noindent The decentralised nature of heating means that data on consumption is not readily available. Unlike electricity, heat does not need to be monitored at high time-resolution to maintain system stability, and the prohibitive cost of heat meters means they are not becoming widespread as are electric smart meters.  This lack of data is a key gap for energy systems modellers, as the difficulty of decarbonising heat, and possible synergies between flexible heating load and intermittent renewable generation rise up the research agenda \citep{huppmann2018new}. \\

\noindent This research seeks to support future studies on energy and climate change mitigation by proposing a new conceptual framework to improve the accuracy and ease with which country-wise heat demand can be estimated based on underlying weather parameters. The novelty of this research is the combined effect of using gas and electricity demand profiles along with weather based data for estimating the country-wise unique heating threshold temperatures along with determining the heating seasons. The focus lays on space heating demand (as opposed to water heating and cooking), as space heating is the majority of final energy demand, and is the one which depends on external conditions such as weather. \\

\noindent In the literature, studies most commonly assume an identical threshold temperature when estimating the heat demand for multiple countries. Heat Roadmap Europe \citep{HRE4} adapts results from  Eurostat \citep{spinoni2015european, eurostat2} where the heating threshold temperature is 18~$\degree$C if the outside temperature drops below 15~$\degree$C. Stratego on the other hand uses 16~$\degree$C for five EU countries \citep{stratego2}. Odyssee uses 18~$\degree$C \citep{bosseboeuf2009energy}. IEA uses 65 $\degree F = 18.3333~\degree$C \citep{iea}. Stratego defines, furthermore, heating seasons differently for the five nations while Odyssee defines a common heating season from October to April for all nations. \\

\noindent A considerable amount of literature has been published on estimating the energy consumption for space heating, using a diverse range of methods. Amongst these, \cite{guo2018machine} use machine learning techniques for time ahead energy demand prediction for building heating systems. \cite{jazizadeh2018personalized} propose a novel approach for which RGB video cameras are used as sensors for measuring personalized thermo-regulation states which can be used as indicators of thermal comfort. \cite{ghahramani2018towards} introduce a hidden Markov model (HMM) based learning method along with infrared thermography of the human face in an attempt to capture personal thermal comfort. \cite{niemierko2019d} use a D-vine copula method to capture the building heating needs by using historical data on German household heating consumption and the respective building parameters. A Modelica library was introduced by \cite{bunning2017modelica} in an attempt to build a control system of building energy systems. \cite{gaitani2010using} use principal component and cluster analysis to create an energy classification tool in an attempt to asses energy savings in different buildings. 
%To the best of our knowledge, no studies in the literature have used a quantitative approach to determine the threshold temperatures. On the other hand, a few available studies have used a qualitative approach to determine the threshold temperatures. In the Smart Controls and Thermal Comfort project (SCATs) \cite{nicol2001scats}, 26 offices in five European countries were visited monthly for heating comfort surveys. These results were utilized by \cite{nicol2007maximum} and concluded that there is no temperature at which everyone would fell comfortable. 
Several studies have also explored the use of weather-based data for estimating heat or gas demand profiles, which is a well recognized practice dating back several decades \citep{aras2008forecasting, timmer2007relations}. It has been applied to multiple case studies as, e.g. \citep{goncu2013forecasting, sarak2003degree} for gas demand estimation or \citep{berger2018novel} for heat demand estimation. We add to the literature by proposing a new approach on how to estimate country-wise comfort temperatures and heating seasons by using historical weather and consumption data. \\

\noindent Two primary data sources are adapted to make this study possible: 1. temperature profiles from a global reanalysis weather model, and 2. data on the national gross consumption of electricity and gas for each country. The choice of data is first of all reflected by the amount of gas and electricity, 43\% and 12\%, respectively, of the final energy demand that is used for heating purposes for the EU \citep{HRE4}. Secondly, the availability of granular data on the consumption makes this study possible. Few of the European countries cover the majority of their heat demand by other fuels such as coal or oil products, as shown in Fig. \ref{fig: fuel_share}. For these fuels, granular data is not available. Further restrictions are introduced by the gas consumption profiles as these are only available for all countries with a monthly resolution and not separated into heating and non-heating sub sectors. Therefore, as a best proxy for the gas consumed in space heating processes the difference in the total gas consumption and gas consumed for electricity generation is explored. Gas consumed for power generation is significant, because gas prices vary from summer to winter relative to coal, and electricity demand increases in winter. Process heating is to a high degree weather independent and thereby not significant for the approximation. Hot water demand and cooking are as well weather independent and so insignificant for this study. Electricity consumption profiles, on the other hand, are typically available with hourly resolution at country level. 

\subsection{Research structure}
\noindent Initially, in the methods section the theory of degree-days is presented and followed by a model for space heat modelling. This is followed by a method to estimate the national-wise heating threshold temperatures for space heating. Next, the Akaike's Information Criteria is presented and used to determine the summer season for each nation. Finally, the temperature validation procedure is presented. The methods section is followed by a description of energy data that is applied in this work. At the end, the results and discussions section along with the conclusion are presented. Additional information is available in the supplementary information.

\section{Nomenclature}

\begin{tabular}{|p{2cm}p{5.5cm}|}
\hline
\textbf{Subscripts} & \textbf{Explanatory text} \\ \hline
$\Delta$ & Time period. \\
$x$ & Grid location. \\
$X$ & Set of grid locations $x$ within a country. \\
$m$ & Model. \\ 
$M$ & Ensemble of models $m$.\\
{} & {} \\
\textbf{Variables} & \textbf{Explanatory text} \\ \hline
$\text{HDD}_{\Delta,X}$ & Heating degree-days for a time period $\Delta$ and country $X$. \\
$T_{0,X}$ & Heating threshold temperature for a country $X$.\\
$T(t)$ & Ambient temperature as a function of time $t$. \\
$p_x$ & Population for a grid location $x$. \\
$p_X$ & Population for a country $X$. \\
$L^\text{heat}$ & Total heat demand. \\
$L^\text{space heat}$ & Space heat demand. \\
$L^\text{hot water}$ & Hot water demand. \\
$L_{0,X}^\text{space heat}$ & Space heat demand per heating degree-day per capita for a country $X$.\\
$\Theta_X$ & Binary indicator function of summer and winter months for a country $X$. \\
$y_{X,\Delta}$ & Measured fuel consumption for a country $X$ and time period $\Delta$. \\ \hline
\end{tabular}

\begin{tabular}{|p{2cm}p{5.5cm}|}
\hline
\textbf{Variables} & \textbf{Explanatory text} \\ \hline
$\hat{y}_{X,\Delta}$ & Modelled fuel consumption for a country $X$ and time period $\Delta$. \\
$\beta_{0,X}$, $\beta_{1,X}$ & Fitting parameters for a country $X$. \\
$\text{AIC}_{m,X}$ & AIC value for a model $m$ and country $X$. \\
$L_{m,X}$ & Likelihood for a model $m$ and country $X$. \\
$F_m$ & Degrees of freedom for a model $m$. \\
$n_m$ & Amount of data points for a model $m$. \\
$\sigma_{X,m}$ & Maximum likelihood estimator for a linear regression for a model $m$ and country $X$.\\
$w_{m,X}$ & AIC weight for a model $m$ and country $X$. \\
$T_x^{adj}$ & Bias adjusted temperature profiles for a grid location $x$.\\
$a_{0,x}$, $a_{1,x}$ & Fitting parameters for a grid location $x$.\\ \hline
\end{tabular}
\section{Methodology}

\subsection{The degree-day method}

\noindent The demand for space heating can be related to the outside ambient temperature by means of the heating degree-day method \citep{thom1954rational,quayle1980heating}, as explained in the following text. \\

\begin{figure*}
    \centering
        \includegraphics[width=0.99\textwidth]{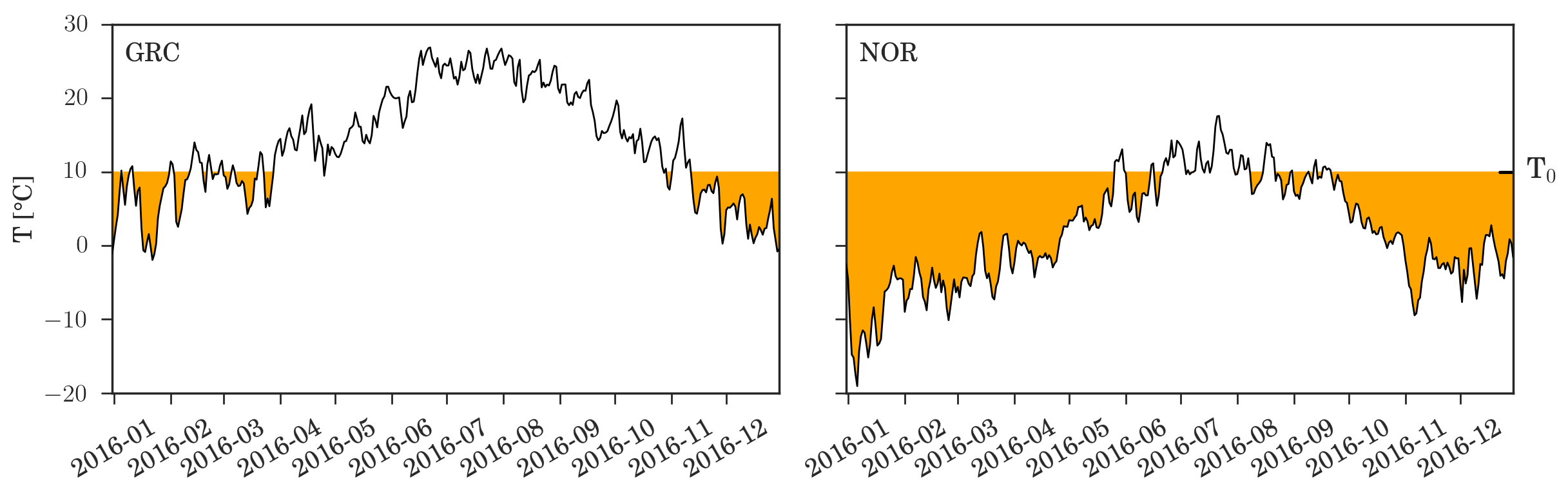}
    \caption{Daily averaged temperatures for Greece and Norway in 2016. The yellow filled area represents the amount of heating degree-days for a heating threshold temperature of 10$\degree$C.} 
    \label{fig: HDD_temp_ex}
\end{figure*}

\noindent Heating degree-days (HDD) are calculated as the integral of the positive difference between a threshold temperature, $T_0$, and the daily average outside temperature, $T$, as illustrated in Fig. \ref{fig: HDD_temp_ex} for Norway and Greece. It is clear that Norway exhibits more HDDs due to its high latitudes, where temperatures are lower during the year. Greece, on the other hand, holds longer summer periods with no HDDs. \\ 

%-------------
 %Figure OK. GBA.
%-------------

\noindent The accumulated heating degree-days, $\text{HDD}_{\Delta,x}$, for a time period, $\Delta$, (e.g. a single day, a week or a month) and grid location, $x$, are related to the threshold temperature as: 

\begin{equation} \label{eq: HDD}
	\text{HDD}_{\Delta,x} = \int_{\Delta} (T_{0,x} - T_x(t))_{+} \; \mathrm{d}t
\end{equation}

\noindent It is assumed that a single threshold temperature, $T_{0,x}$, is used for all grid locations in the set of grid locations, $X$, within a country. The choice of $T_{0,X}$ is not unique and can be chosen according to the region or study \cite{thom1954rational}. $T_x(t)$ denotes the time dependent bias corrected temperature profiles (see Section \ref{sec: bias_correction}). In Eq. \ref{eq: HDD}, $(T_0 - T(t))_{+}$ is defined positive or zero as:

\begin{align*}
(T_{0_x} - T_x(t))_{+}=
\begin{cases}
T_{0,x} - T_x(t) &\text{if} \quad T_{0,x} > T_x(t)\\
0 &\text{if} \quad  T_{0,x} \leq T_x(t)\\
\end{cases}
\end{align*}

\noindent  It is assumed that individual people have the same desire for heating, and so heat demand is proportional to the population density. The national population-weighted heating degree-days, $\text{HDD}_{\Delta,X}$, are then calculated as:

\begin{equation}
	\mathrm{HDD}_{\Delta,\,X} = \frac{1}{p_X} \sum_{x\in X} p_x \cdot \mathrm{HDD}_{\Delta,\,x}
\end{equation}

\noindent where $p_x$ and $p_X$ denote the gridded and total population of a country, respectively. \\

\subsection{Space heat modelling}
\noindent The total heat demand, $L^{\text{heat}}$, is the sum of the demand for space heating, $L^{\text{space heat}}$, and hot water use, $L^{\text{hot water}}$, as shown in Eq. \ref{eq: heat_demand}. 

\begin{equation}
L^{\text{heat}} = L^{\text{space heat}} + L^{\text{hot water}}
\label{eq: heat_demand}
\end{equation}

\noindent Hot water consumption is generally constant throughout the year \citep{staffell2015domestic}, and so assumed to be independent of the ambient temperature. Therefore it is not treated further in this paper.  $L^{\text{space heat}}$, on the other hand, is assumed to be linearly dependent on the HDDs. In the literature, the energy demand for space heating is generally considered to be proportional to the HDDs as in \citep{spinoni2015european, christenson2006climate, berger2018novel}. For a country and a time period, it takes a form as:

\begin{equation} \label{eq: space_heat} 
	L_{\Delta,\,X}^{\text{space heat}} = p_X \cdot L^{\text{space heat}}_{0,X} \cdot \mathrm{HDD}_{\Delta,\,X}\cdot \Theta_X
\end{equation}

\noindent where, $L^{\text{space heat}}_{0,X}$ is a constant equal to the average space heating demand per capita per degree-day in the country $X$. $\Theta_X \in [0,1]$ is a binary indicator function that separates winter from summer months. $\Theta_X=1$ represents winter months where space heating is required. Summer months are represented as $\Theta_X=0$, for which space heating is not required.\\

\noindent In the following section a method to determine the threshold temperature, $T_{0,X}$, is presented and followed by a method to determine the heating season, $\Theta_X$. \\

\subsection{Threshold temperatures for space heating}
\label{sec: space_heat}
\begin{figure}
    \centering
        \includegraphics[width=0.99\columnwidth]{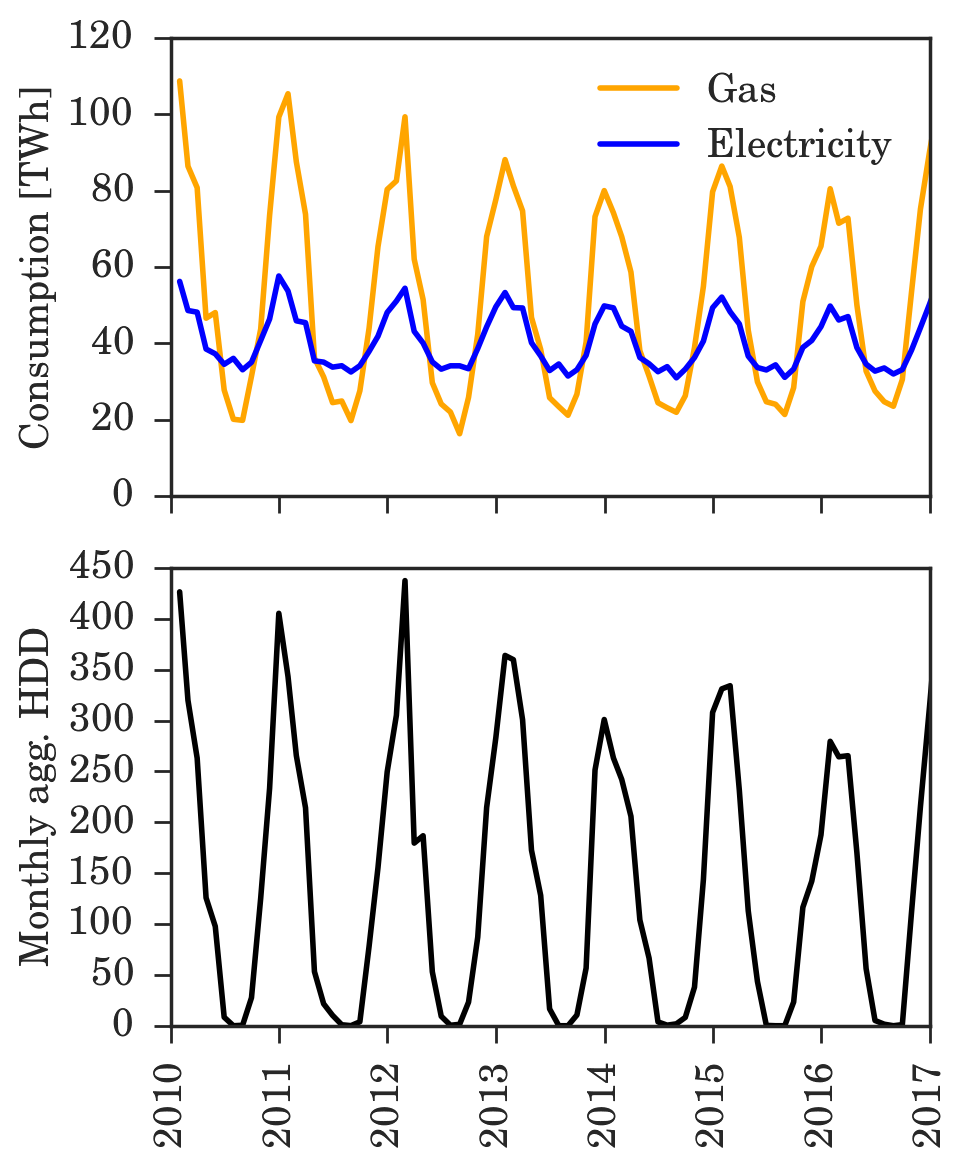}
    \caption{Monthly aggregated gas consumption (orange), electricity consumption (blue) and heating degree-days (black) for France during 2010--2017. The heating degree-days are calculated by using a space heating threshold temperature of 15~$\degree$C.} 
    \label{fig: FRA_ex_gas_el_HDD} 
\end{figure}

\noindent Ideally, the threshold temperature should be determined by comparing HDDs directly to a corresponding time series of heat demand. Since heat demand data is not widely available, except for cities with well monitored district heating networks \citep{dahl2017decision}, the following analysis is based on country aggregated time series data for gas $L^{gas}_X$ or for electricity $L^{el}_X$ consumption. This method can also be applied to actual heat demand data. Data on gas and electricity consumption is available for all European countries, as described in Section~\ref{section:data}, as well as many other countries. An example of gas and electricity consumption along with monthly aggregated HDDs for France is shown in Fig. \ref{fig: FRA_ex_gas_el_HDD}. \\

\noindent Gas and electricity are typically converted using boilers, resistance heaters or heat pumps, which operate with comparable efficiency over the year (given that most heat pumps in Europe are ground source rather than air source) \citep{staffell2012review}. Referring to Eq.~\ref{eq: space_heat}, this motivates the following model for either the gas or the electricity consumption time series as a function of HDDs for a country, $X$:

\begin{equation}\label{eq:OnlyWinterModel}
	\hat{y}_{X,\Delta}(t;T_{0,X}) = \beta_{0,X} \cdot \text{HDD}_{X,\Delta}(t;T_{0,X}) + \beta_{1,X}
\end{equation}

\noindent where $\hat{y}_{X,\Delta}(t;T_{0,X})$ is the modelled consumption, i.e. gas or electricity, summed over the period, $\Delta$, and evaluated at time, $t$. $\beta_{0,X}$ and $\beta_{1,X}$ are model parameters that are assumed to be independent of both time and temperature. $\beta_{1,X}$ defines consumption of gas or electricity that cover all domestic energy demand apart from space heating. Space heating is introduced through $\beta_{0,X}$. Finally, only $\text{HDD}_{X,\Delta}$ is assumed to depend on $T_0$. Note that this relation only applies for winter months where $\Theta_X=1$.\\

\noindent The model parameters, $\beta_{0,X}$ and $\beta_{1,X},$ as well as the best choice of $T_{0,X}$ for a country are determined by minimizing the root mean square of the errors between the modelled consumption, $\hat{y}_{X,\Delta}$, of gas and electricity and the corresponding measured consumption, $y_{X,\Delta}$ as:

\begin{align}
\min\limits_{T_{0,X},\beta_{0,X},\beta_{1,X}} \quad & RMSE  = \sqrt{\frac{1}{n}\sum_{t}\left(\hat{y}_{X,\Delta}(t;T_{0,X})-y_{X,\Delta}(t)\right)^2}  \nonumber  \\
\text{s.t.} \quad &5 \leq T_{0,X} \leq 25 \label{eq: opt_th}
\end{align}

\noindent where $n$ is the sample size. In the following analysis, independent values for $T_{0,X}$ are calculated for each year of data, and then a median is taken to calculate a single value along with the 25th to 75th percentile significance range. The optimal values of $\beta_{0,X}$ and $\beta_{1,X}$  relate to the energy mix and population of a country and are not discussed further in this study.\\

%-------------
%-------------

\subsection{Heating seasons}
\label{sec: heating_season}
\noindent During summer months when space heating is turned off, both gas and electricity demands are assumed to be independent of HDDs. A constant summer demand for a country, $\beta_{1,X}$, is the simplest model that describes this relationship. Thus, Eq.~\ref{eq:OnlyWinterModel} is extended in the following way:

\begin{flalign}
&\text{Winter model:} \nonumber \\
&\hat{y}_{X,\Delta}(t;T_{0,X}) = \beta_{0,X} \cdot HDD_{X,\Delta}(t;T_{0,X}) + \beta_{1,X} \\ 
&\nonumber \\
&\text{Summer model:} \nonumber \\
&\hat{y}_{X,\Delta}(t) = \beta_{1,X}
\end{flalign}

\noindent To make a self-consistent determination of the model parameters and of the classification of the data into summer or winter months, an initial guess is undertaken in which the three warmest months: June, July and August are used to determine the single free parameter $\beta_{1,X}$ of the summer model and November-March the resulting winter model free parameters $T_{0,X}$ and $\beta_{0,X}$. Then, all months are classified by using Akaike's Information Criterion (AIC), as described below, and the winter and summer model parameters are recalculated by using this classification. This process may be repeated until model parameters and classification reach convergence, which usually happened after a single repetition.\\

\noindent The Akaike's Information Criterion, AIC \citep{akaike1987factor} is a well-recognized procedure for model selection, which takes both descriptive accuracy and parsimony into account. The objective of the AIC model selection is to quantify the information lost when the probability distribution associated with a model is used to represent the probability distribution of the data. The classification is then performed by choosing the model with the lowest expected information loss, and, thus, the lowest AIC value \citep{akaike1987factor}. The AIC for a model $m$ is defined as:  

\begin{equation}
\text{AIC}_{m,X} = -2\log(L^{\max}_{m,X}) + 2F_m + \frac{2F_m\left(F_m+1\right)}{n-F_m-1}
\label{eq: AIC}
\end{equation}

\noindent $L^{\max}_{m,X}$ represents the maximum likelihood value for a model and country, while $F_m$ represents the degrees of freedom for a model. $n_m$ represents the amount of data points for a model. $L_{m,X}$ is shown in Eq. \ref{eq:max_like}. $\sigma^2_{m,X}$, as shown in Eq. \ref{eq:max_like_sig}, is the maximum likelihood estimator that is country and model specific. The maximization of $L_{m,X}$ rewards the accuracy, leading to lower AIC while more free parameters penalizes the lack of parsimony and leading to higher AICs. The third term in Eq. \ref{eq: AIC} is a modification \citep{hurvich1995model}, which is recommended if $\frac{n_m}{F_m} < 40$ \citep{burnham2003model}. 

\begin{equation}
L_{m,X} = \left(2\pi\sigma^2_{m,X}\right)^{-\frac{n}{2}}\exp^{-\dfrac{1}{2\sigma^2_{m,X}}\sum_{t}\left(\hat{y}_{m,X,\Delta}(t)-y_{X,\Delta}(t)\right)^2}
\label{eq:max_like}
\end{equation}

\begin{equation}
\sigma^2_{m,X} = \frac{1}{n} \left(\sum_{t}\left(\hat{y}_{m,X,\Delta}(t)-y_{X,\Delta}(t)\right)^2\right)
\label{eq:max_like_sig}
\end{equation}

\noindent It is also important to address the weight of evidence of choosing the model with the lowest AIC. The Akaike weight, $w_{m,X}\left(\text{AIC}\right)$, \citep{burnham2003model} is defined as:

\begin{equation}
w_{m,X}\left(\text{AIC}_{m,X} \right) = \frac{\exp^{-\frac{1}{2}\Delta_{m,X}\text{AIC}}}{\sum_{m=1}^M \exp^{-\frac{1}{2}\Delta_{m,X}\text{AIC}}}
\end{equation}

\noindent where $\sum_{m,X} w_{m,X}\left(\text{AIC}_{m,X}\right) = 1$. $\Delta_{m,X} \text{AIC} = \text{AIC}_{m,X} - \min\left(\text{AIC}_{M,X}\right)$ and $M$ denotes the ensemble of possible models. \\

\noindent In general, a preferred model is accepted if the evidence ratio exceeds 2 \citep{anderson1998comparison}, alternatively, an ensemble average of models is recommended. In this work, the AICs are respected regardless of the evidence ratio but in cases of a low evidence ratio extra attention is paid to the classification. These issues arise mostly in Spring and Autumn, where the outdoor temperatures vary significantly.\\ 

\begin{figure*}
    \centering
        \includegraphics[width=0.99\textwidth]{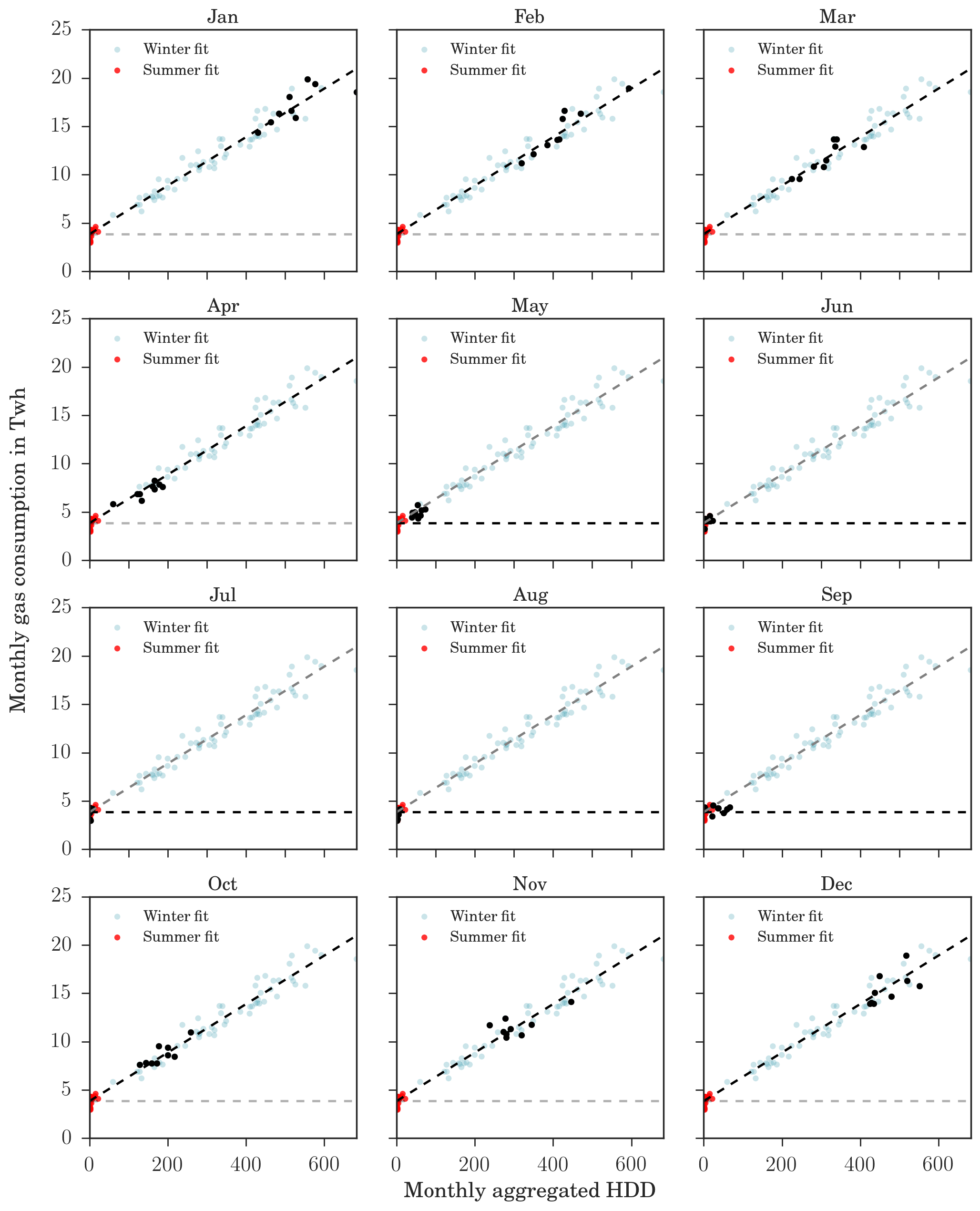}
    \caption{Winter (blue) and summer (red) classes with monthly data (red) to be classified for Hungary spanning the years 2009-2018. The winter class is trained by the blue coloured data while the summer class is trained by the red coloured data.} 
    \label{fig: HUN_ex}
\end{figure*}

\noindent Fig. \ref{fig: HUN_ex} exemplifies the method, described above, for the case of using gas for heating in Hungary. Test data belonging to each month (shown with black) is classified into either of the two classes. January to April are classified as Winter months with high evidence ratios as seen in Tab. \ref{tab: AIC_ex}. May is classified as a summer month but with a very weak evidence ratio and the model selection is indecisive. June to September are classified as Summer months. October to December are classified as winter months with strong evidence ratios. The classification of September shows the importance of a summer model. For this month the national gas consumption show no significant relation to the heating degree-days and so gas is not used for heating purposes. \\

\begin{table*}
\centering
\caption{AIC values of the classification for Hungary in the case for heating by gas.}
\begin{tabular}{lllllllllllll}
HUN & Jan & Feb & Mar & Apr & May & Jun  \\
Winter: & 52.04 (1.0)  & 47.82 (1.0) & 45.35 (1.0) & 41.89 (1.0) & 36.08 (0.43) & 34.21 (0.02)  \\
Summer: & 82.58 (0.0) & 78.39 (0.0) & 72.53 (0.0) & 56.54 (0.0) & 35.54 (0.57) & 26.72 (0.98)  \\
&&&&&&\\
{} & Jul & Aug & Sep & Oct & Nov & Dec\\
Winter: & 34.37 (0.02) & 31.45 (0.01) & 39.36 (0.06) & 41.18 (1.0) & 43.08 (1.0) & 47.08 (1.0)\\
Summer: & 26.64 (0.98)  & 22.90 (0.99) & 33.97 (0.94)  & 62.14 (0.0) & 71.46 (0.0) & 79.47 (0.0)\\
\label{tab: AIC_ex}
\end{tabular}
\end{table*}

%-------------
%-------------

\subsection{Bias adjustment of temperature profiles}
\label{sec: bias_correction}
\noindent This section presents a simple method that can be used to bias correct any gridded temperature profiles. In this study, a best representation of real temperature data is important for a correct estimation of the HDDs.  \\

\noindent The reanalysis ground temperature data comes from the Climate Forecast System Reanalysis (CFSR) data set, which is supplied by the National Center for Atmospheric Research (NCAR) \citep{saha2010ncep}. This data covers the entire globe from 1979 to present and is updated on a monthly basis. The spatial and temporal resolution covers Europe with $0.312\degree$ and 1~hour, respectively. The main advantage of using global reanalysis data is a high availability in all locations, consistency across many decades, and preservation of correlations between different weather fields relevant for energy system analysis, e.g. temperature, wind, solar and precipitation data.  However, it should be used with caution as local biases may be larger compared to, e.g., data from mesoscale models or ground measurements. \\

\noindent Temperature data based on direct measurements comes from the European Climate Assessment (ECAD) who provide an interpolation in space \citep{haylock2008european}. The data set covers Europe for the period 1950--2017 with a spatial and temporal resolution of 0.5$\degree$ and 1~day, respectively. The underlying measurement data covers various time periods depending on the mast operation span. In a simple bias correction procedure the CFSR reanalysis temperature data was compared to the ECAD ground measurements in all grid locations, $x$, for a daily temporal resolution. ECAD ground measurements were interpolated in space by the "nearest neighbour" method to meet the resolution of CFSR. \\ 

\begin{figure}
	\centering	
	\includegraphics[width=0.99\columnwidth, height=0.95\columnwidth]{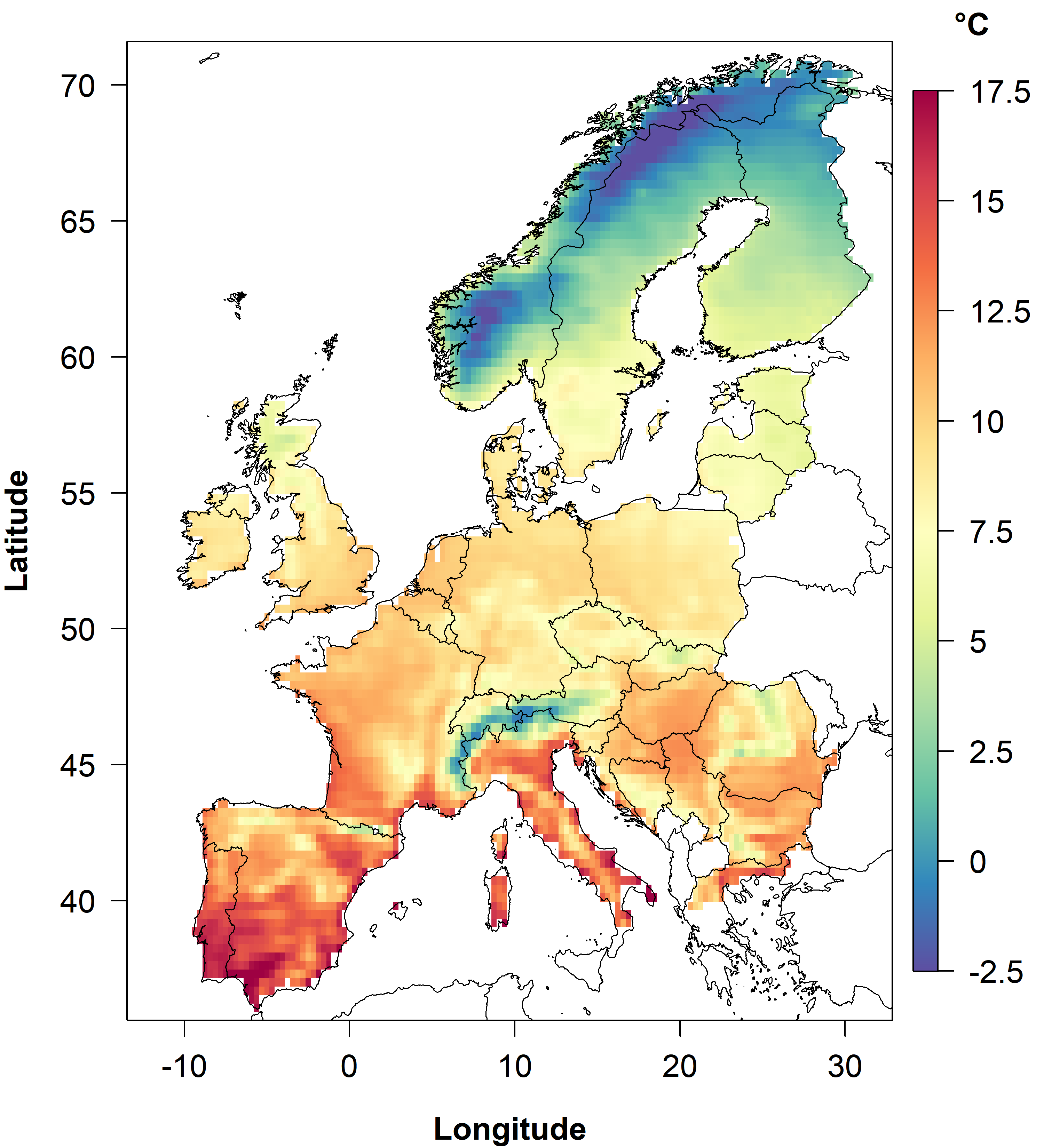}	
	\includegraphics[width=0.99\columnwidth, height=0.95\columnwidth]{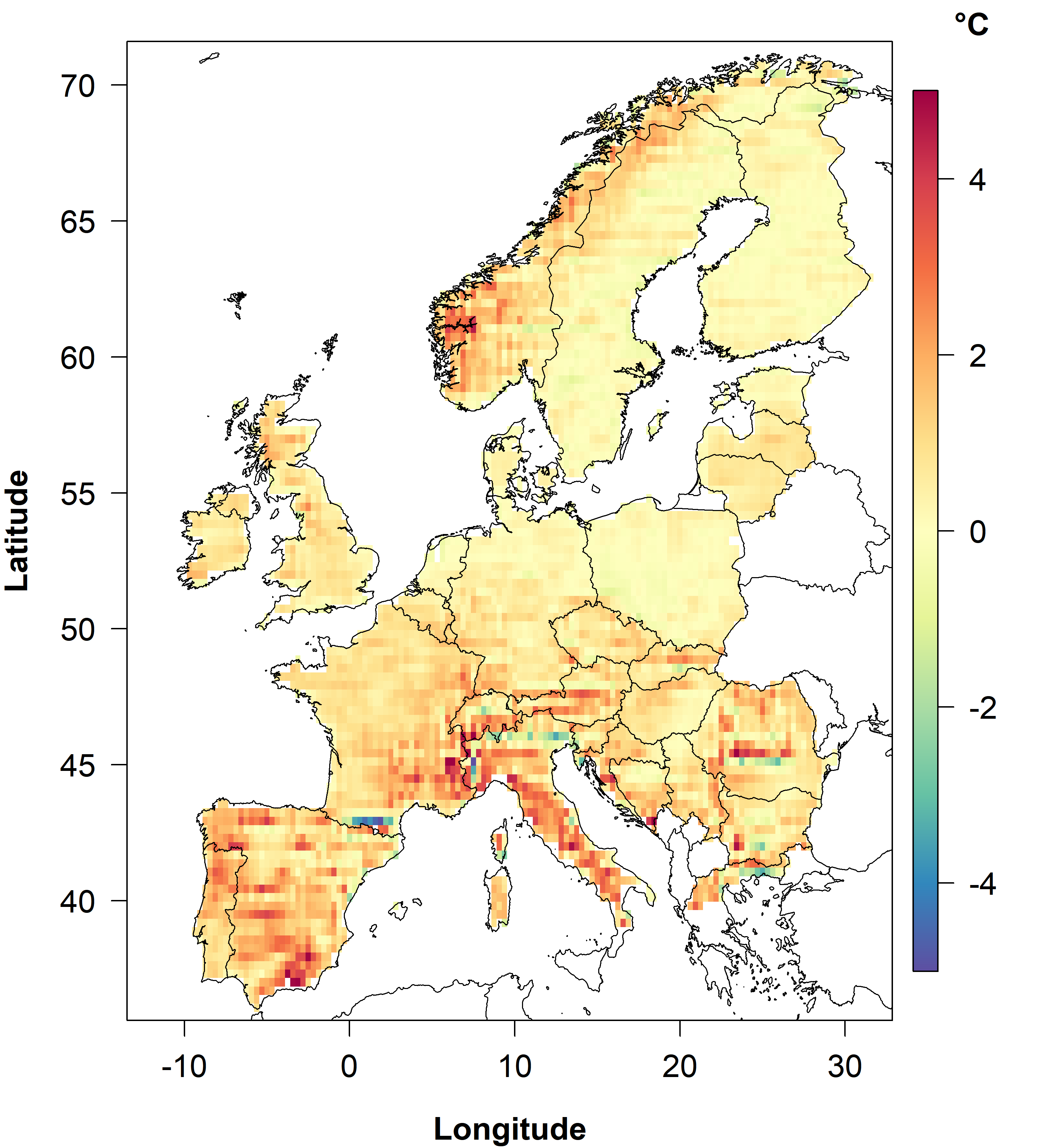}	
	\caption{Upper plot: Spatial distribution of the uncorrected average temperatures from 2005--2010. Lower plot: Spatial distribution of the average temperature correction.}
	\label{fig:corr_facts} 
\end{figure}

\noindent In a linear regression, as in Eq. \ref{Eq: lin_reg}, the ECAD temperature data set acts as a predictor variable while the CFSR temperature data set acts as the response variable. The least square estimators $\alpha_0$ and $\alpha_1$ denote, as usual, the gradients and offsets, respectively. The system of linear equations is solved for every grid cell, $x$, contained within the set of grid cells $X$. The bias adjusted CFSR temperature profiles are finally calculated as:

\begin{equation}
T^{adj}_x(t) = \frac{1}{\alpha_{0,x}} T_x(t) - \frac{\alpha_{1,x}}{\alpha_{0,x}}
\label{Eq: lin_reg}
\end{equation}

\noindent where $t \in [0;1826]$ denotes the day number in the period from 01/01/2011 to 31/12/2015. The corrections are summarized in Fig. \ref{fig:corr_facts}. In general, the corrections are relavily small, and the most extreme bias correction parameters are observed in sparsely populated mountainous regions as, e.g. the Alps, Sierra Nevada, Sierra Blanca and the West chain of the Norwegian mountains. \\

%-------------
%-------------

\section{Energy data}
\label{section:data}

\noindent Acquisition of energy consumption data varies significantly between energy sectors and countries, and nationwide data on energy use with high granularity are generally not available. This data gap introduces a serious weakness for energy system research. In the following detailed information is provided on the data that is used in this study. \\

\subsection{Electricity consumption data}
\label{sec: el_data}
\noindent National electricity consumption profiles with hourly resolution were acquired from the European Network of Transmission System Operators for Electricity, (ENTSO-E). The data covers the period 2006 -- 2017 \citep{entsoe}. Data from 2009 and earlier is limited to the member TSOs of the Continental Europe region. Data from 2010 and on includes all ENTSO-E members. National data for the UK \citep{staffell2018increasing}, France \citep{bossmann2015shape} and Denmark \citep{energinet} were obtained seperatly to correct for gaps and inconsistencies in the ENTSO-e data.

\subsection{Gas consumption data}
\label{sec: gas_data}
\noindent Data on electricity production from gas is available through the ENTSO-E transparency platform with daily resolution \citep{entsoe}. From this, the amount of gas that was used to produce electricity is estimated with a conversion efficiency of 51.5\%. Total national gas consumption with monthly resolution is available through Eurostat from 2008 to 2018 \citep{eurostat}. This covers all end-uses, including consumption by the gas sector it self, but excludes export. End use consumption includes the residential, service, industrial and agriculture sectors. Data on gas entering and exiting a country is metered by the national gas TSOs with a daily resolution and made available through the ENTSO-G transparency platform from earliest September 2013 \citep{entsog}. National gas consumption with daily resolution is then estimated for a few countries by the difference in the amount entering and existing gas. The UK national gas consumption excluding the share of gas used in electricity production was provided by the UK TNO. Danish total gas consumption was provided by Energinet \citep{energinet}. \\

%To complement gas demand as a proxy for heat demand, for Denmark data on heat production by district heating utilities are provided on time scales from 3 minutes to one hour. \\

\section{Results and Discussions}
\noindent Initially, we present results from a first analysis where the iterative procedure (described in Section \ref{sec: heating_season}) has been used to estimate the threshold temperatures and the corresponding heating seasons for all countries. For this, exclusively monthly aggregated gas and electricity consumption data were used. In both cases, the iteration converged after the second cycle. Next, we adapt the resulting classification and recalculate the threshold temperatures by using weekly and daily gas and electricity consumption data. This allows for an assessment of the influence of data granularity. \\

\begin{figure*}[!bt]
    \centering
        \includegraphics[width=0.99\textwidth]{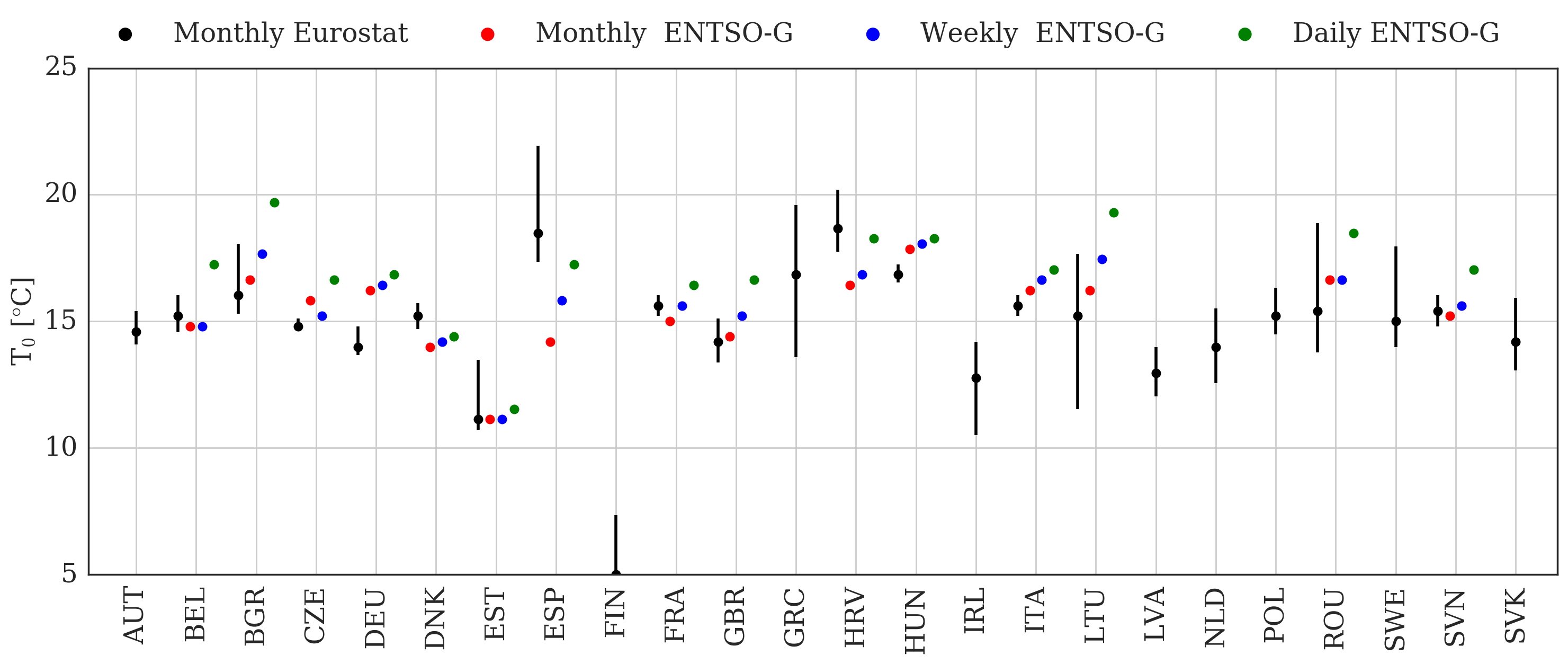}
    \caption{Median of yearly heating threshold temperatures for the monthly eurostat gas data (black) and monthly (red), weekly (blue) and daily (green) ENTSO-G gas data. Threshold temperatures for Denmark and UK were recalculated by using data from national sources explained in Section \ref{sec: gas_data} and showed with red, blue and green colors. $\left[ q_{25\%},q_{75\%}\right]$ uncertainty range is provided for the monthly eurostat gas consumption data. Switzerland, Serbia and Bosnia \& Herzegovina are not shown due to missing data. Norway and Portugal are not shown as heating by gas is classified as non-existing for these countries.}
    \label{fig: gas_all}
\end{figure*}

\begin{figure*}[!bt]
    \centering
        \includegraphics[width=0.99\textwidth]{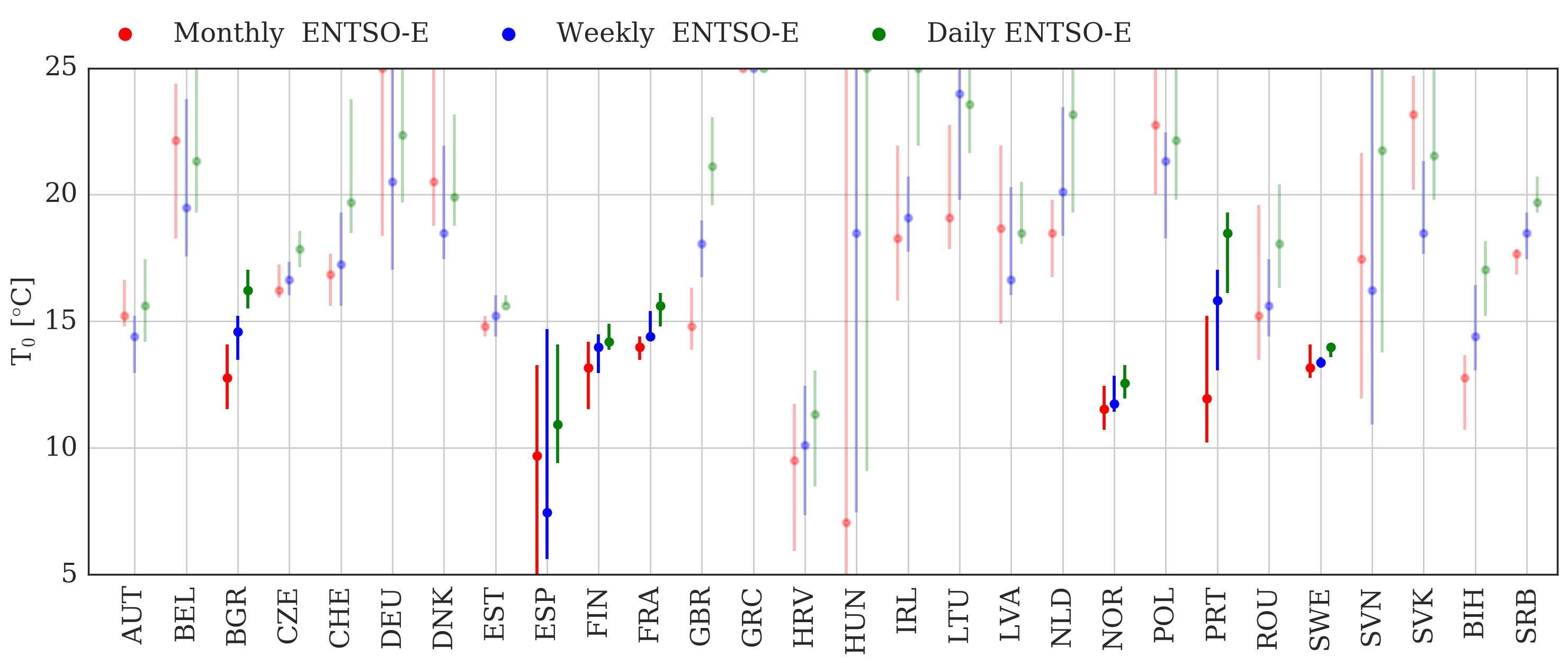}
    \caption{Median of yearly heating threshold temperatures with $\left[ q_{25\%},q_{75\%}\right]$ uncertainty range determined by using electricity consumption data with daily (black), weekly (blue) and monthly (red) resolution. Countries of which the final heat demand is covered by less than 15\% by electricity are shown with faint colors. Results for all countries apart from Denmark, France and UK were obtained by using electricity consumption data provided by ENTSO-E. Results for France, Denmark and UK were obtained by using data from national sources as stated in Section \ref{sec: el_data}. Italy is not shown as heating by electricity is classified as non-existing.}     
    \label{fig: el_all}
\end{figure*}

\noindent Fig. \ref{fig: gas_all} and \ref{fig: el_all} show the median score of the yearly threshold temperatures which were computed by using daily, weekly and monthly aggregated gas and electricity consumption data, respectively. Nine yearly values of the threshold temperature allows for a determination of the corresponding $\left[ q_{25\%},q_{75\%}\right]$ uncertainty ranges for the monthly Eurostat gas consumption data and ENTSO-E electricity consumption data. In the following, we only focus on results computed by using monthly aggregated consumption data. \\

\noindent From Fig. \ref{fig: gas_all} it is clear that the estimated threshold temperatures by using Eurostat (black) and ENTSO-G (red) gas consumption data  are not significantly different within the Eurostat uncertainty range. Threshold temperatures for Norway and Portugal are not shown as by classification no space heating demand is covered by gas. For Norway, this is in agreement with radical changes in the Norwegian energy system with a ban of using gas for domestic heating by 2020. Results for Spain, Greece, Lithuania and Romania appear with substantial 25th to 75th percentile uncertainties. These are not unexpected as for these countries, gas covers a minor share of the final energy demand (Fig. \ref{fig: fuel_share}) and, consequently, no penetrative relation might be developed to the weather. There are, however, other possible explanations as, e.g., data quality or quantity. In the case of heating by electricity, a majority of the countries show unstable threshold temperatures along with extensive 25th to 75th percentile range. As for heating by gas, these results could have impacts from several sources. A few countries as Finland, France, Norway and Sweden show valid based on small error scores. \\

\noindent Results based on monthly consumption data have been summarized in Tab. \ref{tab: ht_summary}. Results are not presented where a fuel type covers less than 15\% of the final heating demand, as below this, the relationship between fuel consumption and heating degree-days (Eq. \ref{eq:OnlyWinterModel}) lost statistical significance. A few countries hold a heating threshold temperature for both fuel types. It is clear that threshold temperatures for heating by electricity are smaller in comparison to heating by gas. It is difficult to explain this result, but it might be related to that electricity is a more expensive source of heating in countries for which gas is the predominantly heating source. Therefore, electricity could be used as a supplementary for gas during extreme temperature drops. In general, the ensemble of country-wise threshold temperatures for heating by gas average to 15.0$\pm$1.7 $\degree$C (1 sigma standard deviation). The electricity values average to 13.4$\pm$2.4 $\degree$C. \\

%-------------
%-------------

%-------------
%-------------

\begin{table}
\centering
\caption{Heating threshold temperatures for heating by gas and electricity with uncertainty ranges. n.a denotes a share of fuel type below 15\% and results are not trusted.}
\footnotesize
\renewcommand{\arraystretch}{1.3}
\begin{tabular}{|ccc|} \hline
{} & \textbf{Electricity} & \textbf{Gas - Eurostat}\\
Country & T$_{0} \left[q_{25\%},q_{75\%}\right]$ $\degree$C & T$_{0} \left[q_{25\%},q_{75\%}\right]$ $\degree$C\\ \hline
AUT &  n.a. & 14.59             $\left[ 14.08,15.41 \right]$ \\
BEL &  n.a. & 15.20             $\left[ 14.59,16.02 \right]$ \\
BGR &  12.76 $\left[ 11.53,14.08 \right]$ & 16.02 $\left[ 15.31,18.06 \right]$ \\
CZE &  n.a. & 14.80 $\left[ 14.80,15.10 \right]$ \\
CHE &  16.84 $\left[ 15.61,17.65 \right]$ & n.a.\\
DEU &  n.a. & 13.98 $\left[ 13.67,14.80 \right]$ \\
DNK &  n.a. & 15.20 $\left[ 14.69,15.71 \right]$ \\
EST &  n.a. & 11.12 $\left[ 10.71,13.47 \right]$ \\
ESP &  9.69 $\left[ 5.00,13.27 \right]$ & 18.47 $\left[ 17.35,21.94 \right]$ \\
FIN &  13.16 $\left[ 11.53,14.18 \right]$ & n.a. \\
FRA &  13.98 $\left[ 13.47,14.39 \right]$  & 15.61 $\left[ 15.20,16.02 \right]$ \\
GBR &  n.a.  & 14.18 $\left[ 13.37,15.10 \right]$ \\
GRC &  n.a. & 16.84 $\left[ 13.57,19.59 \right]$ \\
HRV &  n.a.  & 18.67 $\left[ 17.76,20.20 \right]$ \\
HUN &  n.a.  & 16.84 $\left[ 16.53,17.24 \right]$ \\
IRL &  n.a.  & 12.76 $\left[ 10.51,14.18 \right]$ \\
ITA &  n.a.  & 15.61 $\left[ 15.20,16.02 \right]$ \\
LTU &  n.a.  & 15.20 $\left[ 11.53,17.65 \right]$ \\
LVA &  n.a.  & 12.96 $\left[ 12.04,13.98 \right]$ \\
NLD &  n.a.  & 13.98 $\left[ 12.55,15.51 \right]$ \\
NOR &  11.53 $\left[ 10.71,12.45 \right]$ & n.a.  \\
POL &  n.a.  & 15.2 $\left[ 14.49,16.33 \right]$ \\
PRT &  11.94 $\left[ 10.20,15.20 \right]$ & n.a.  \\
ROU &  n.a. & 15.41 $\left[ 13.78,18.88 \right]$ \\
SWE &  13.16 $\left[ 12.76,14.08 \right]$ & n.a.  \\
SVN &  n.a.  & 15.41 $\left[ 14.80,16.02 \right]$ \\
SVK &  n.a.  & 14.18 $\left[ 13.06,15.92 \right]$ \\
BIH &  12.76 $\left[ 10.71,13.67 \right]$ & n.a.  \\
SRB &  17.65 $\left[ 16.84,17.86 \right]$ & n.a.  \\ \hline
\end{tabular}
\renewcommand{\arraystretch}{1}
\label{tab: ht_summary}
\end{table}

\noindent Tab. \ref{tab: month_classification} presents a 10 year average (2008-2017) of monthly aggregated heating degree-days for each country. Enveloped months represent the summer season for which space heating is usually not required, since the heat absorbed during daylight hours is enough to keep the buildings warm during colder periods. The binary indicator function, $\Theta _X$, takes a values of zero for the enveloped months and one for the rest.  Countries for which threshold temperatures are available for both heating by gas and electricity, the minimum required heating season is shown. It is clear that all countries exhibit a summer period from June-August. Apart from this, the classification shows a spread in the summer months, which mostly depends on the geographical position of the countries. As could be expected, South European countries usually hold longer summer periods without heating while the Northern countries tend to have shorter summer periods. \\

\noindent Daily and weekly aggregated gas and electricity consumption data belonging to the winter classified months have been used to recalculate the  heating threshold temperatures. The results are shown in Fig. \ref{fig: gas_all} and \ref{fig: el_all}, respectively. For both consumption types, the statistical similarity in threshold temperatures for each individual country provide a robust indication of the adequacy of using less granular data for estimating the threshold temperatures. On the other hand, it is clear that the threshold temperatures increase with increasing data granularity. \\

\noindent In the following we illustrate the significance of reaching country specific heating threshold temperatures and summer seasons. As a case study, results for Great Britain are used but an identical analysis can be performed for each individual country by utilizing the heating degree-days in Tab. \ref{tab: month_classification}. For Great Britain, October averages to 90 heating degree-days, and is classified as a winter month, while May, which as well averages to 90 heating degree-days, is not. Contrary to May, the winter classified October is explained by an existing relation between gas consumption and heating degree-days. On the other hand, the AIC evidence ratio for May is below 2 and, thus, more years of consumption data would be needed to fully justify this classification. A similar classification is shown for Hungary for May and September as observed in Fig. \ref{fig: HUN_ex}. Similar cases appear for Denmark, Estonia, Greece, Romania, Switzerland and Bosnia \& Herzegovina as shown by Tab. \ref{tab: month_classification}. These issues arise mostly during Autumn and Spring where the monthly temperature differences exhibit large variances over the years.  \\

\begin{figure}
    \centering
        \includegraphics[width=0.48\textwidth]{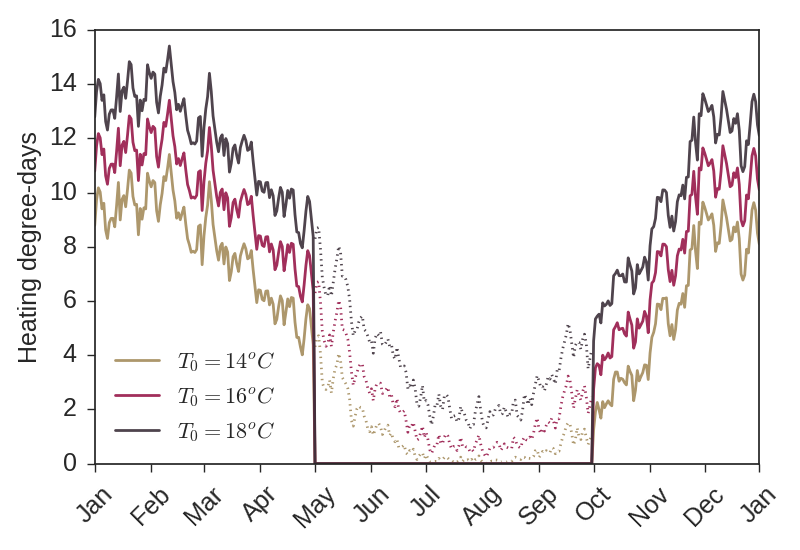}
    \caption{Average of 10 yearly heating degree-days for Great Britain calculated with heating threshold temperatures, $T_0 = 14\degree$C (yellow), $T_0 = 16\degree$C (red) and $T_0 = 18\degree$C (black). Fully drawn lines illustrate the heating degree-days during winter months as a result of the classification. Dotted lines illustrate summer months for which space heating is not needed and have to be removed.}     
    \label{fig: HDDvsMonths}
\end{figure}

\noindent  Average heating degree-days calculated by using a threshold temperature of 14 $\degree$C, 16 $\degree$C and 18 $\degree$C for Great Britain are shown in Fig. \ref{fig: HDDvsMonths}. The summer season is shown by a depreciation of heating degree-days from May to October. It is clear that a 2 $\degree$C increase in the threshold temperature introduce a significant difference in the accumulated heating degree-days over a year. The most striking result which emerges from the classification is the extreme change in the seasonal pattern of the heating degree-days. \\

%-------------
%-------------

\noindent Quantitative measures of the heating degree-days are shown in Tab. \ref{tab: HDDvsMonths_summary} for six case studies. In the most extreme scenario, case study c) overestimates the heating degree-days by approximately 93\%, which is almost a doubling in comparison to case study d). For a fixed average space heat demand per capita per heating degree-day, $L^\text{space heat}_{0,\text{GBR}}$, the energy demand for space heating, Eq. \ref{eq: space_heat}, is consequently overestimated by identical shares. These results suggest that the current estimations of the energy demand for space heating in various projects might be highly over or underestimated for some countries. This might introduce further changes as, e.g., the estimation of CO$_2$ emissions, technology choice for heating or peak demand estimation. On the other hand, a yearly fixed energy consumption for space heating will be distributed differently according to the seasonal distribution of heating degree-days.\\

\begin{table}
\centering
\caption{Overview of yearly aggregated heating degree-days for six case studies of Great Britain denoted by a)-f).}
\footnotesize
\renewcommand{\arraystretch}{1.3}
\begin{tabular}{|c|cc|} \hline
{} & \textbf{With summer season} & \textbf{Without summer season}\\ \hline
$\boldsymbol{T_0 = 14\degree}$\textbf{C} & a) 1654 & d) 1510\\
$\boldsymbol{T_0 = 16\degree}$\textbf{C} & b) 2235 & e) 1927\\
$\boldsymbol{T_0 = 18\degree}$\textbf{C} & c) 2896 & f) 2350  \\ \hline
\end{tabular}
\renewcommand{\arraystretch}{1}
\label{tab: HDDvsMonths_summary}
\end{table}

\begin{figure}
    \centering
        \includegraphics[width=0.5\textwidth]{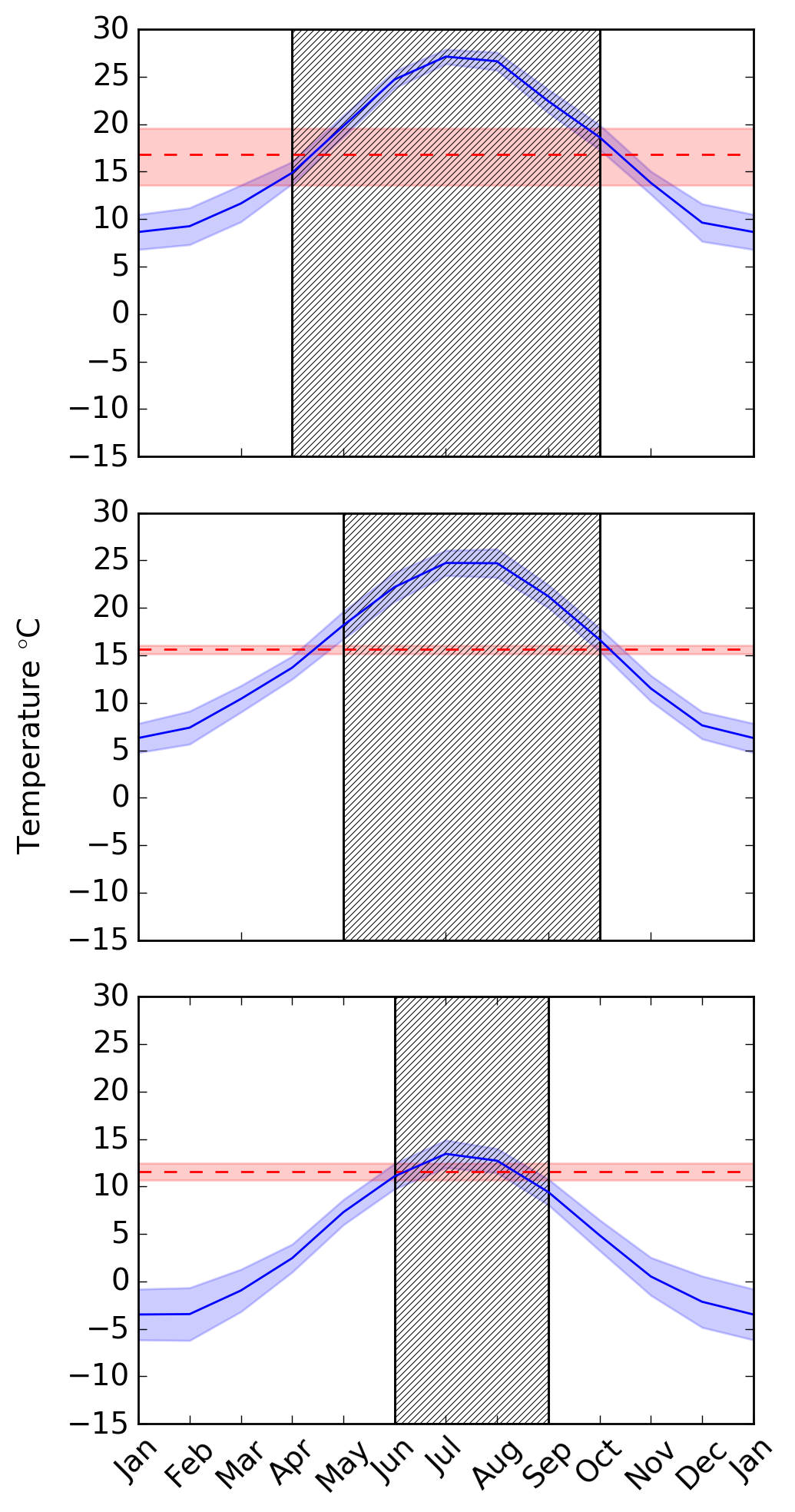}
    \caption{Monthly averaged temperatures from 2008-2017  with one sigma uncertainty range (blue full drawn curve with shaded region), heating threshold temperature with $\left[q_{25\%},q_{75\%}\right]$ uncertainty range (red dashed line with shaded region) and classified summer season (black hatched area) for Greece (upper figure), Italy (central figure) and Norway (lower figure).}     
    \label{fig: val}
\end{figure}

\noindent Fig. \ref{fig: val} illustrate the synergy between the monthly averaged ground  temperature measurements (blue curve), the threshold temperature (red dashed line) and the classified summer season (hatched area) for Greece, Italy and Norway. From these figures it is clear that the monthly averaged temperature falls below the heating threshold temperature outside the hatched area, which indicates that space heating is needed.\\

%\noindent Energy demand for space heating will be reduced in every country for a corresponding decrease in the heating degree-days. By using Eq. \ref{eq: space_heat} we find an average of 15\% or 40 TWh/yr reduction in space heat demand for Great Britain from 2008 to 2017 from omitting the heating degree-days that are present during the Summer period. Here we assume a population of $p_{\text{GBR}} = $ 64.4 m and $L^{\text{space heat}}_{0,\text{GBR}}=$ 320kJ/cap/HDD for the residential and commercial sectors. $T_0=$14.03$\degree$C and $\Theta_{\text{GBR}}$ from Tab. \ref{tab: month_classification}. An identical heating threshold temperature as in Eurostat and HRE provides a heat demand of 474TWh/yr which is approximately double our finding. By using 16$\degree$C and 18$\degree$C as in Stratego and Odyssee or EIA we find 353TWh/yr and 458TWh/yr.

\subsection{Country wise validation}

\noindent \textbf{Denmark} has no heating season defined by law. 54 \% of the end-use heat demand is provided by district heating \citep{nordic} which dominates the Danish heat production. During summer time the district heating utilities mainly deliver hot water. A similar summer season is determined in this work with a heating threshold temperature of 15.20 $\left[ 14.69,15.71 \right]$ $\degree$C. A similar finding was presented by \cite{dahl2017decision}. \\%From Fig. \ref{fig: hvidesande}-\ref{fig: skagen} it is clear that the Danish district heating utilities stop providing heat from June to September. \\

\noindent \textbf{Czech Republic} has a legally defined heating season that lasts from September 1st to May 31st \citep{stratego2}. If the daily average outside temperature is below 13$\degree$C then the district heating utilities start to deliver heat. An identical heating season is determined by this work with a threshold temperature of 14.80 $\left[ 14.80,15.10 \right]$ $\degree$C. A possible explanation for this discrepancy is that our results covers the complete heating production by electricity and gas while 13$\degree$C only refers to district heating. \\

\noindent \textbf{Great Britain} has no heating season provided by law but a typical heating season starts by October 1st and ends at April 30th and is further restricted with a day time peak temperature being 16 $\degree$C or lower for a few consecutive days \citep{stratego2}. An identical heating season is proposed in this study with a threshold temperature of 14.18 $\left[ 13.37,15.10 \right]$ $\degree$C which is conductive with 16$\degree$C daytime and 9$\degree$C night temperatures. A threshold temperature of $13\degree$C is proposed by the \cite{HSE} based on qualitative surveys.\\

\noindent \textbf{Germany} has no legal heating season, but the German Tenants Association, \citep{DMB}, states the heating season typically runs from October 1st to April 30th. This gives a heating season that is two months shorter than found here, which is explained by the threshold temperatures. The Association of German engineers, VDI 2067, estimate a German heating threshold temperature of 12$\degree$C, whereas a threshold of 13.98 $\left[ 13.67,14.80 \right]$ is presented by this study. \\

\noindent \textbf{Finland} also has no heating season defined by law. According to \cite{jylha2015hourly} an accepted heating threshold temperature is 12$\degree$C from Autumn to December and lowers to 10$\degree$C during the Spring. Here, a value of 13.16 $\left[ 11.53,14.18 \right]$ $\degree$C is proposed to be used from September through to May. \\

\noindent \textbf{Italy} has several heating seasons defined by law depending on six different climatic zones from the mountainous North with a colder climate to the flat South with a temperate climate \citep{stratego2}. October 15th is the earliest date at which heating is permitted and lasts at most to April 15th. A national-aggregate heating period is found to run from October 1st to March 31st with a threshold of 15.61 $\left[ 15.20,16.02 \right]$ $\degree $C. \\

\noindent \textbf{Croatia} has a typical heating season to range from September 15th to May 15th. The heating season is in this study proposed to start at September 1st and last to April 30th with a heating threshold temperature of 18.67 $\left[ 17.76,20.20 \right]$ $\degree $C. \\

\noindent \textbf{Romania}'s district heating utilities begin to operate by law if the outside average temperature reaches 10$\degree $C or lower for three consecutive days, and no later than November 1st. Heat delivery stops, by law, if the daily average temperature exceeds 10$\degree $C for three consecutive days and not earlier than April 15th. In this work, the overall heating season is found to start from October 1st and last to March 31st with a heating threshold temperature of 15.41 $\left[ 13.78,18.88 \right]$ $\degree $C. \\

\noindent \textbf{Spain} holds heating threshold temperatures from 13-14.8 $\degree$C depending on the region \citep{labandeira2012estimation, blazquez2013residential}. In this work, 9.69 $\left[ 5.00,13.27 \right]$ $\degree$C is found for electricity use and 18.47 $\left[ 17.35,21.94 \right]$ $\degree$C for gas use.

%-------------
%-------------

\section{Conclusion}
\noindent This study was undertaken to design a new method to determine the national energy demand for residential and commercial space heating with better than annual resolution. Furthermore, the study was designed to work for any given country based on its historic fuel consumption and weather data. In doing so, we propose a new method to determine a consistent empirically-derived national-wise heating threshold temperature which can be used to determine the national aggregated heating degree-days. Secondly, the extent of a winter period is determined for which space heating is required. This is represented by a binary indicator function taking only values of zeros and ones. The final energy demand for space heating is then a function of the newly acquired heating degree-days and the binary indications. As a case study, these methods have been applied to the majority of European countries, using data on national aggregated heating degree-days along with national demand for gas and electricity. The following conclusions can be drawn:

\begin{enumerate}
\item The ensemble mean of country specific heating threshold temperatures for heating by gas average to 15.0$\pm$1.7 $\degree$C (1 sigma standard deviation). The electricity values average to 13.4$\pm$2.4 $\degree$C. This suggests that the currently defined threshold temperatures might be overestimated by up to 5$\degree$C for some countries in the literature.\\
\item The heating threshold temperatures were computed by using daily, weekly and monthly aggregated data and shown to be marginally different within 25th to 75th percentile significance range. This provides a good indication of the adequacy of using monthly aggregated data for such an application. \\
\item The heating threshold temperatures tend to increase with increasing temporal resolution of consumption and weather data. \\
\item It was also shown that European countries exhibit a summer period of at least June, July and August where space heating is not required. South European countries exhibit longer summer periods, up to a maximum of 9 months in Portugal.
\end{enumerate}

\noindent Ultimately, we find hetereogenous threshold temperatures for space heating across neighbouring countries in Europe, which suggest that the threshold temperatures cannot be extrapolated to neighbouring countries. And so, this study excels from the standard practice of using a blanket, even arbitrary value, such as 16$\degree$C or 18$\degree$C across multiple countries which over-simplify and misrepresent the true nature and scale of national space heating demand. If an unrepresentative heating threshold temperature is used then the seasonal behaviour of the heat demand becomes incorrect. In fact, a further in-depth study of Great Britain suggest that current estimations of the energy demand for space heating might be overestimated by 93\% in comparison to the findings of this work. This might introduce further changes in the heating technology choices or the peak demand estimations.\\

\noindent Some limitations of this study need to be considered. Firstly, the lack of data for national gas consumption specifically for space heating purposes adds a layer of complexity, as the end-uses of space heating, water heating, cooking and other industrial processes cannot be disentangled.  Gas consumed for electricity generation could be estimated, but consumption data would also be useful for this purpose.  ENTSO-G and other authorities could consider increasing the transparency of data reporting to aid further research on heating demand.  Secondly, this study did not evaluate data on coal and oil consumption, which was not found at monthly or better resolution.  This may prove important for countries such as Poland and Greece, and would further enhance results. Thirdly, countries with various climatic areas and a diverse terrain might hold different threshold temperatures in different regions, as might countries with large socio-economic differences between regions. Assessing sub-regions of countries would require more granular data on electricity and gas demand than is currently made openly available, but would be an interesting future topic for research.\\

\section*{Acknowledgement}
\noindent Thanks to Aarhus University Research Foundation for funding S. Kozarcanin with funding number AUFF-E-2015-FLS-7-26. I. Staffell acknowledges the Engineering and Physical Sciences Research Council for funding the IDLES project (EP/R045518/1). G. B. Andresen was funded by the RE-INVEST project, which is supported by the Innovation Fund Denmark under grant number 6154-00022B.

\section*{Declaration of interest}
\noindent Declarations of interest: none

\section*{Correspondence} 
\noindent Correspondence should be addressed to S. Kozarcanin (email: sko@eng.au.dk) or G. B. Andresen (email: gba@eng.au.dk). 

\begin{table*}
\centering
\caption{Results of the classification of summer months by the AIC model selection criteria. Each number denotes the monthly averaged heating degree-days from 2008 to 2017 using the combined temperature thresholds as shown in Tab. \ref{tab: ht_summary}. Enveloped values denotes classified summer months.}
\scriptsize
\renewcommand{\arraystretch}{1.5}
\begin{tabular}{|ccccccccccccc|}
\hline
Country  & Jan & Feb & Mar & Apr & May & Jun & Jul & Aug & Sep & Oct & Nov & Dec \\ \cline{7-9}
AUT & $\color{Gray}{482}$ & $\color{Gray}{406}$ & $\color{Gray}{315}$ & $\color{Gray}{171}$ & $\color{Gray}{78}$ & \multicolumn{1}{|c}{$\color{black}{26}$} & $\color{black}{13}$ & \multicolumn{1}{c|}{$\color{black}{15}$} & $\color{Gray}{62}$ & $\color{Gray}{181}$ & $\color{Gray}{287}$ & $\color{Gray}{431}$ \\ \cline{6-6} \cline{10-10}
BEL & $\color{Gray}{366}$ & $\color{Gray}{317}$ & $\color{Gray}{264}$ & $\color{Gray}{161}$ & \multicolumn{1}{|c}{$\color{black}{75}$} & $\color{black}{19}$ & $\color{black}{4}$ & $\color{black}{4}$ & \multicolumn{1}{c|}{$\color{black}{33}$} & $\color{Gray}{119}$ & $\color{Gray}{224}$ & $\color{Gray}{328}$ \\\cline{5-5}
BIH & $\color{Gray}{380}$ & $\color{Gray}{309}$ & $\color{Gray}{226}$ & \multicolumn{1}{|c}{$\color{black}{95}$} & $\color{black}{33}$ & $\color{black}{4}$ & $\color{black}{1}$ & $\color{black}{0}$ & \multicolumn{1}{c|}{$\color{black}{22}$} & $\color{Gray}{96}$ & $\color{Gray}{180}$ & $\color{Gray}{333}$ \\ \cline{5-5}
BGR & $\color{Gray}{502}$ & $\color{Gray}{386}$ & $\color{Gray}{304}$ & $\color{Gray}{163}$ & \multicolumn{1}{|c}{$\color{black}{53}$} & $\color{black}{8}$ & $\color{black}{1}$ & $\color{black}{1}$ & \multicolumn{1}{c|}{$\color{black}{28}$} & $\color{Gray}{147}$ & $\color{Gray}{250}$ & $\color{Gray}{422}$ \\ \cline{5-5} \cline{10-10}
CHE & $\color{Gray}{556}$ & $\color{Gray}{496}$ & $\color{Gray}{426}$ & \multicolumn{1}{|c}{$\color{black}{292}$} & $\color{black}{188}$ & $\color{black}{92}$ & $\color{black}{59}$ & \multicolumn{1}{c|}{$\color{black}{65}$} & $\color{Gray}{153}$ & $\color{Gray}{272}$ & $\color{Gray}{390}$ & $\color{Gray}{520}$  \\ \cline{5-6}
CZE & $\color{Gray}{503}$ & $\color{Gray}{416}$ & $\color{Gray}{325}$ & $\color{Gray}{173}$ & $\color{Gray}{71}$ & \multicolumn{1}{|c}{$\color{black}{16}$} & $\color{black}{3}$ & \multicolumn{1}{c|}{$\color{black}{5}$} & $\color{Gray}{51}$ & $\color{Gray}{187}$ & $\color{Gray}{292}$ & $\color{Gray}{437}$ \\ \cline{10-10}
DEU & $\color{Gray}{401}$ & $\color{Gray}{344}$ & $\color{Gray}{274}$ & $\color{Gray}{147}$ & $\color{Gray}{58}$ & \multicolumn{1}{|c}{$\color{black}{12}$} & $\color{black}{1}$ & $\color{black}{2}$ & \multicolumn{1}{c|}{$\color{black}{31}$} & $\color{Gray}{134}$ & $\color{Gray}{241}$ & $\color{Gray}{351}$ \\ \cline{6-6}
DNK & $\color{Gray}{427}$ & $\color{Gray}{393}$ & $\color{Gray}{355}$ & $\color{Gray}{229}$ & \multicolumn{1}{|c}{$\color{black}{110}$} & $\color{black}{29}$ & $\color{black}{2}$ & $\color{black}{3}$ & \multicolumn{1}{c|}{$\color{black}{34}$} & $\color{Gray}{147}$ & $\color{Gray}{255}$ & $\color{Gray}{365}$ \\ 
EST & $\color{Gray}{474}$ & $\color{Gray}{416}$ & $\color{Gray}{372}$ & $\color{Gray}{206}$ & \multicolumn{1}{|c}{$\color{black}{56}$} & $\color{black}{5}$ & $\color{black}{0}$ & $\color{black}{0}$ & \multicolumn{1}{c|}{$\color{black}{13}$} & $\color{Gray}{140}$ & $\color{Gray}{241}$ & $\color{Gray}{363}$ \\ \cline{5-5}\cline{11-11}
ESP & $\color{Gray}{175}$ & $\color{Gray}{152}$ & $\color{Gray}{106}$ & \multicolumn{1}{|c}{$\color{black}{46}$} & $\color{black}{14}$ & $\color{black}{1}$ & $\color{black}{0}$ & $\color{black}{0}$ & $\color{black}{1}$ & \multicolumn{1}{c|}{$\color{black}{13}$} & $\color{Gray}{78}$ & $\color{Gray}{156}$ \\  \cline{5-6} \cline{10-11}
FIN & $\color{Gray}{620}$ & $\color{Gray}{536}$ & $\color{Gray}{497}$ & $\color{Gray}{326}$ & $\color{Gray}{136}$ & \multicolumn{1}{|c}{$\color{black}{37}$} & $\color{black}{4}$ & \multicolumn{1}{c|}{$\color{black}{11}$} & $\color{Gray}{74}$ & $\color{Gray}{264}$ & $\color{Gray}{368}$ & $\color{Gray}{507}$ \\ 
FRA & $\color{Gray}{356}$ & $\color{Gray}{311}$ & $\color{Gray}{247}$ & $\color{Gray}{150}$ & $\color{Gray}{70}$ & \multicolumn{1}{|c}{$\color{black}{16}$} & $\color{black}{3}$ & \multicolumn{1}{c|}{$\color{black}{4}$} & $\color{Gray}{28}$ & $\color{Gray}{104}$ & $\color{Gray}{211}$ & $\color{Gray}{320}$ \\ \cline{6-6} \cline{10-10}
GBR & $\color{Gray}{301}$ & $\color{Gray}{270}$ & $\color{Gray}{248}$ & $\color{Gray}{171}$ & \multicolumn{1}{|c}{$\color{black}{90}$} & $\color{black}{26}$ & $\color{black}{5}$ & $\color{black}{5}$ & \multicolumn{1}{c|}{$\color{black}{26}$} & $\color{Gray}{90}$ & $\color{Gray}{193}$ & $\color{Gray}{270}$ \\ \cline{5-5}
GRC & $\color{Gray}{287}$ & $\color{Gray}{234}$ & $\color{Gray}{194}$ & \multicolumn{1}{|c}{$\color{black}{92}$} & $\color{black}{17}$ & $\color{black}{0}$ & $\color{black}{0}$ & $\color{black}{0}$ & \multicolumn{1}{c|}{$\color{black}{3}$} & $\color{Gray}{41}$ & $\color{Gray}{112}$ & $\color{Gray}{236}$ \\ \cline{5-5}
HRV & $\color{Gray}{523}$ & $\color{Gray}{438}$ & $\color{Gray}{349}$ & $\color{Gray}{202}$ & \multicolumn{1}{|c}{$\color{black}{100}$} & $\color{black}{24}$ & $\color{black}{6}$ & $\color{black}{6}$ & \multicolumn{1}{c|}{$\color{black}{72}$} & $\color{Gray}{209}$ & $\color{Gray}{318}$ & $\color{Gray}{471}$ \\ 
HUN & $\color{Gray}{521}$ & $\color{Gray}{414}$ & $\color{Gray}{309}$ & $\color{Gray}{144}$ & \multicolumn{1}{|c}{$\color{black}{50}$} & $\color{black}{7}$ & $\color{black}{1}$ & $\color{black}{1}$ & \multicolumn{1}{c|}{$\color{black}{39}$} & $\color{Gray}{178}$ & $\color{Gray}{302}$ & $\color{Gray}{465}$ \\ \cline{5-5} \cline{11-11}
IRL & $\color{Gray}{225}$ & $\color{Gray}{204}$ & $\color{Gray}{197}$ & \multicolumn{1}{|c}{$\color{black}{135}$} & $\color{black}{65}$ & $\color{black}{15}$ & $\color{black}{2}$ & $\color{black}{2}$ & $\color{black}{16}$ & \multicolumn{1}{c|}{$\color{black}{62}$} & $\color{Gray}{152}$ & $\color{Gray}{207}$ \\ \cline{5-5} \cline{11-11}
ITA & $\color{Gray}{287}$ & $\color{Gray}{241}$ & $\color{Gray}{173}$ & $\color{Gray}{74}$ & \multicolumn{1}{|c}{$\color{black}{22}$} & $\color{black}{4}$ & $\color{black}{2}$ & $\color{black}{2}$ & \multicolumn{1}{c|}{$\color{black}{10}$} & $\color{Gray}{49}$ & $\color{Gray}{134}$ & $\color{Gray}{259}$ \\ 
LTU & $\color{Gray}{595}$ & $\color{Gray}{493}$ & $\color{Gray}{428}$ & $\color{Gray}{245}$ & \multicolumn{1}{|c}{$\color{black}{87}$} & $\color{black}{25}$ & $\color{black}{2}$ & $\color{black}{7}$ & \multicolumn{1}{c|}{$\color{black}{68}$} & $\color{Gray}{254}$ & $\color{Gray}{340}$ & $\color{Gray}{480}$ \\
LVA & $\color{Gray}{519}$ & $\color{Gray}{436}$ & $\color{Gray}{375}$ & $\color{Gray}{199}$ & \multicolumn{1}{|c}{$\color{black}{55}$} & $\color{black}{8}$ & $\color{black}{0}$ & $\color{black}{0}$ & \multicolumn{1}{c|}{$\color{black}{26}$} & $\color{Gray}{184}$ & $\color{Gray}{274}$ & $\color{Gray}{401}$ \\
NLD & $\color{Gray}{335}$ & $\color{Gray}{299}$ & $\color{Gray}{247}$ & $\color{Gray}{141}$ & \multicolumn{1}{|c}{$\color{black}{61}$} & $\color{black}{11}$ & $\color{black}{0}$ & $\color{black}{1}$ & \multicolumn{1}{c|}{$\color{black}{14}$} & $\color{Gray}{88}$ & $\color{Gray}{192}$ & $\color{Gray}{290}$ \\\cline{10-10} \cline{6-6}
NOR & $\color{Gray}{440}$ & $\color{Gray}{393}$ & $\color{Gray}{350}$ & $\color{Gray}{221}$ & $\color{Gray}{107}$ & \multicolumn{1}{|c}{$\color{black}{33}$} & $\color{black}{8}$ & \multicolumn{1}{c|}{$\color{black}{11}$} & $\color{Gray}{47}$ & $\color{Gray}{172}$ & $\color{Gray}{281}$ & $\color{Gray}{391}$ \\
POL & $\color{Gray}{521}$ & $\color{Gray}{434}$ & $\color{Gray}{362}$ & $\color{Gray}{200}$ & $\color{Gray}{83}$ & \multicolumn{1}{|c}{$\color{black}{22}$} & $\color{black}{4}$ & \multicolumn{1}{c|}{$\color{black}{7}$} & $\color{Gray}{61}$ & $\color{Gray}{208}$ & $\color{Gray}{306}$ & $\color{Gray}{444}$ \\\cline{5-6} \cline{10-11}
PRT & $\color{Gray}{73}$ & $\color{Gray}{62}$ & $\color{Gray}{31}$ & \multicolumn{1}{|c}{$\color{black}{10}$} & $\color{black}{1}$ & $\color{black}{0}$ & $\color{black}{0}$ & $\color{black}{0}$ & $\color{black}{0}$ & \multicolumn{1}{c|}{$\color{black}{0}$} & $\color{Gray}{23}$ & $\color{Gray}{64}$ \\ \cline{11-11}
ROU & $\color{Gray}{530}$ & $\color{Gray}{417}$ & $\color{Gray}{305}$ & \multicolumn{1}{|c}{$\color{black}{140}$} & $\color{black}{42}$ & $\color{black}{6}$ & $\color{black}{1}$ & $\color{black}{2}$ & \multicolumn{1}{c|}{$\color{black}{32}$} & $\color{Gray}{161}$ & $\color{Gray}{275}$ & $\color{Gray}{451}$ \\\cline{5-5}
SRB & $\color{Gray}{507}$ & $\color{Gray}{400}$ & $\color{Gray}{300}$ & $\color{Gray}{144}$ & \multicolumn{1}{|c}{$\color{black}{55}$} & $\color{black}{10}$ & $\color{black}{2}$ & $\color{black}{2}$ & \multicolumn{1}{c|}{$\color{black}{37}$} & $\color{Gray}{164}$ & $\color{Gray}{279}$ & $\color{Gray}{448}$ \\
SVK & $\color{Gray}{507}$ & $\color{Gray}{410}$ & $\color{Gray}{317}$ & $\color{Gray}{155}$ & \multicolumn{1}{|c}{$\color{black}{58}$} & $\color{black}{13}$ & $\color{black}{4}$ & $\color{black}{6}$ & \multicolumn{1}{c|}{$\color{black}{48}$} & $\color{Gray}{178}$ & $\color{Gray}{284}$ & $\color{Gray}{447}$ \\ 
SVN & $\color{Gray}{486}$ & $\color{Gray}{411}$ & $\color{Gray}{312}$ & $\color{Gray}{157}$ & \multicolumn{1}{|c}{$\color{black}{63}$} & $\color{black}{11}$ & $\color{black}{3}$ & $\color{black}{4}$ & \multicolumn{1}{c|}{$\color{black}{52}$} & $\color{Gray}{176}$ & $\color{Gray}{282}$ & $\color{Gray}{441}$ \\ \cline{6-6}\cline{10-10}
SWE & $\color{Gray}{500}$ & $\color{Gray}{442}$ & $\color{Gray}{397}$ & $\color{Gray}{248}$ & $\color{Gray}{109}$ & \multicolumn{1}{|c}{$\color{black}{26}$} & $\color{black}{2}$ & \multicolumn{1}{c|}{$\color{black}{7}$} & $\color{Gray}{49}$ & $\color{Gray}{196}$ & $\color{Gray}{304}$ & $\color{Gray}{429}$ \\ \hline
\end{tabular}
\renewcommand{\arraystretch}{1}
\label{tab: month_classification}
\end{table*}

\bibliographystyle{elsarticle-harv} 
\bibliography{bib}

\end{document}